\documentclass[aps,prb,twocolumn,superscriptaddress,showpacs]{revtex4}
\usepackage{amsmath}
\usepackage{graphicx,epsfig}

\newcommand{\be}{\begin{equation}}
\newcommand{\ee}{\end{equation}}
\newcommand{\bary}{\begin{eqnarray}}
\newcommand{\eary}{\end{eqnarray}}

\begin{document}

\title{Lattice-coupled Antiferromagnet on Frustrated Lattices}
\author{Chenglong Jia}
%\email[Electronic address:]{cl_jia@hotmail.com}
\affiliation{Department of Physics, BK21 Physics Research
Division, Sung Kyun Kwan University, Suwon 440-746, Korea}
\author{Jung Hoon Han}
\email[Electronic address:]{hanjh@skku.edu}
\affiliation{Department of Physics, BK21 Physics Research
Division, Sung Kyun Kwan University, Suwon 440-746, Korea}
\affiliation{CSCMR, Seoul National University, Seoul 151-747,
Korea}
\date{\today}

\begin{abstract}
Lattice-coupled antiferromagnetic spin model is analyzed for a
number of frustrated lattices: triangular, Kagom\'{e}, and
pyrochlore. In triangular and Kagom\'{e} lattices where ground
state spins are locally ordered, the spin-lattice interaction does
not lead to a static deformation of the lattice. In the pyrochlore
structure, spin-lattice coupling supports a picture of the hexagon
spin cluster proposed in the recent experiment[S. H. Lee \textit{\
et al.}, Nature, \textbf{418}, 856 (2002)]. Through spin-lattice
interaction a uniform contraction of the individual hexagons in
the pyrochlore lattice can take place and reduce the exchange
energy. Residual hexagon-hexagon interaction takes the form of a
3-states Potts model where the preferred directions of the
spin-loop directors for nearby hexagons are mutually orthogonal.
\end{abstract}
\maketitle

The continued interest in insulating antiferromagnet appears to
have a twofold objective. One is the search for a novel type of
quantum ground state, other than the conventional N\'{e}el-ordered
state, particularly in two dimensions\cite{senthil}. A
resonating-valence-bond ground state, if it should exist, is
believed to lead naturally to a superconducting phase when doped
with holes\cite{anderson}. In a separate development,
understanding the nature of antiferromagnetic ground state defined
on the frustrated lattice, even at the classical level, has been
the focus of much theoretical\cite{theory} as well as experimental
activity\cite{exp}.

In both these cases the fundamental Hamiltonian describing the spin
interaction is
\begin{equation}
H = \sum_{\langle ij \rangle} J_{ij} S_i \cdot S_j ~~ (J_{ij}>0)
\end{equation}
defined for an appropriate set of bonds $\langle ij \rangle$. The
antiferromagnetic interaction for an insulating magnet is mediated by the
superexchange process in which overlap of Wannier orbitals localized at
different positions provides the necessary (virtual) hopping mechanism. As
such, it is not surprising that the exchange energy $J_{ij}$ should depend
on the separation of orbitals. Existing experimental data suggests that $%
J_{ij}$ falls off as 6-14th power of the
separation\cite{exp-on-Jij}.

Writing the equilibrium position of the ions by $i$ and $j$, and
the displacement vector of each ion by $u_i$ and $u_j$, the
exchange integral has the expansion in the small displacements
\begin{equation}
J_{ij}=J(|i+u_i - j-u_j |) \approx J_0 - J_1 \hat{e}_{ji} \cdot (u_j - u_i )
\end{equation}
where $\hat{e}_{ji} = (j-i)/|j-i|$ is the unit vector, and $J_0$
and $J_1$ are positive constants. For an Einstein, or optical,
phonon mode and ignoring the kinetic energy of the displacement,
we arrive at the lattice-coupled spin model\cite{pytte}
\begin{equation}
H = \sum_{\langle ij \rangle} \left(J_0 - J_1 \hat{e}_{ji} \cdot (u_j - u_i
) \right) S_i \cdot S_j + {\frac{K}{2}} \sum_i u_i^2 .
\label{spin-phonon-model}
\end{equation}
The purpose of this paper is to analyze the ground state of the model
Hamiltonian, Eq. (\ref{spin-phonon-model}), for a number of frustrated
lattices: triangular, Kagom\'{e}, and pyrochlore\cite{recent-papers}.

By rescaling the displacements, $u_{i}\rightarrow u_{i}/\sqrt{K}$, $J_{1}$
is rescaled to $J_{1}/\sqrt{K}\equiv \alpha $, while the overall energy
scale is fixed by $J_{0}$, which is set to one. The reduced Hamiltonian has
the form
\begin{equation}
H=\sum_{\langle ij\rangle }S_{i}\cdot S_{j}-\alpha \sum_{i}u_{i}\cdot f_{i}+{%
\frac{1}{2}}\sum_{i}u_{i}^{2}
\end{equation}%
with $f_{i}=\sum_{j\in i}\hat{e}_{ij}S_{i}\cdot S_{j}$. We introduce the
notation $j\in i$ to indicate all bonds $j$ that are exchange-coupled to site $%
i $. Minimizing the energy gives the condition relating the lattice
positions with the spins:
\begin{equation}
u_{i}/\alpha =\langle f_{i}\rangle =\sum_{j\in i}\hat{e}_{ij}\langle
S_{i}\cdot S_{j}\rangle .  \label{u-to-S}
\end{equation}

Classical antiferromagnet on a triangular lattice has a ground state
characterized by a $120^{\circ }$ angle between a pair of adjacent spins.
Long-range order is established at the mean-field level. When Eq. (\ref%
{u-to-S}) is applied to the ground state spin arrangements of the triangular
antiferromagnet, one finds $u_{i}=0$. Unrestricted numerical solution of Eq.(%
\ref{u-to-S}) together with the classical mean-field equations for the spin
average $\langle S_{i}\rangle $ also yields $u_{i}=0$, consistent with a
preliminary Monte Carlo simulation of the model Eq.(\ref{spin-phonon-model})%
\cite{han}. Thus, at the classical level, the ground state of the
classical antiferromagnet on a triangular lattice is unaffected by
the coupling to lattice. Employing the acoustic phonon model,
$(K/2)\sum_{\langle ij\rangle }(u_{i}-u_{j})^{2}$, alters the
mean-field equation to $\sum_{j\in
i}(u_{i}-u_{j})/\alpha =\langle f_{i}\rangle $, which also gives zero for $%
u_{i}$\cite{comment}.

The effective Hamiltonian after \textquotedblleft integrating out" the
displacement $u_{i}$,
\begin{equation}
H_{eff}=\sum_{\langle ij\rangle }S_{i}\cdot S_{j}-{1\over2}{\alpha
^{2} }\sum_{i}f_{i}^{2},  \label{Effective-H}
\end{equation}%
can be expanded around its classical minimum by use of the
Holstein-Primakoff (HP) theory. Up to quadratic order in the HP bosons\cite%
{HP},
\begin{equation}
H_{eff}=E_{0}+\frac{zS}{2}\sum\limits_{k}\left[ \omega _{k}\left(
b_{k}^{+}b_{k}+\frac{1}{2}\right) -\frac{A_{k}}{2}\right]
\end{equation}%
with
\begin{gather*}
A_{k}=1+\frac{\gamma _{k}}{2},\text{ \ \ }B_{k}=\frac{3\gamma _{k}}{4}, \\
E_{0}=-\frac{1}{4}NzS^{2},\text{ \ \ }\omega _{k}=\sqrt{%
A_{k}^{2}-(2B_{k})^{2}},
\end{gather*}%
where $\gamma _{k}=\frac{1}{z}\sum_{\langle ij\rangle }e^{ik\cdot
(r_{j}-r_{i})}$, $z=6$ is the coordination number for triangular
lattice, and $N$ is the number of sites. There are no terms in the
effective Hamiltonian proportional to $\alpha ^{2}$ up to
quadratic order in the HP bosons, and the ``vacuum" of the
Hamiltonian remains unaltered after the introduction of
spin-lattice coupling.

Average of $%
f_{i}$ can be worked out within the quadratic theory, yielding%
\begin{equation}
\langle f_{i}\rangle =\left[
{\frac{S^{2}}{2}}\!-\!\frac{S}{N}\sum_{k}\left[
\omega _{k}\left( n_B (k)\!+\!\frac{1}{2}\right) \!-\!\frac{A_{k}}{2}\right] %
\right] \sum_{j\in i}\hat{e}_{ji}=0
\end{equation}%
for the triangular lattice, and $n_B (k)$ is the boson occupation
number of energy $\omega _{k}$. Therefore, small quantum or
thermal fluctuation fails to produce a lattice distortion, or
\textquotedblleft spin-Peierls effects" as it is often known in
the literature. The higher-order terms in the HP bosons may have
interesting consequences, including a spontaneous distortion of
the lattice, and shall be considered in the future.
%%%%%%%%%%%%%%%%%%%%%%%%%%%%%%%%%%%%%%%%%%%%%%%%%%%%%%%%%%%%%%%%%%%%%%%%%%%%%%%%%
\begin{figure}[b]
\includegraphics[angle=90,width=8cm]{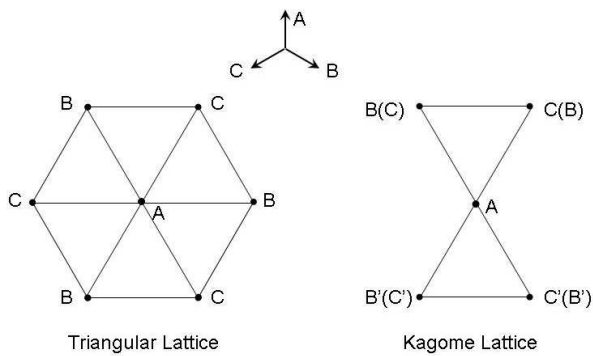}
\caption{{\protect\small Ground-state spin configuration of the
triangular and Kagom\'{e} lattice. $A,B,C$ refer to three
orientations of the spin.}} \label{tri-and-kagome}
\end{figure}
%%%%%%%%%%%%%%%%%%%%%%%%%%%%%%%%%%%%%%%%%%%%%%%%%%%%%%%%%%%%%%%%%%%%%%%%%%%%%%%%%

Classical ground state of the antiferromagnet on the Kagom\'{e}
lattice is defined by the condition $\sum_{i\in \tau }S_{i}=0$
where $i=1,2,3$ are the three corners of a triangle $\tau$, for
all $\tau $. This leads to the well-known ground-state degeneracy
of order $2^{N}$, where $N$ is the number of triangles in the
lattice. Nevertheless, ground-state spin-spin correlation
$S_{i}\cdot S_{j}$ equals $-1/2$ for all nearest-neighbor pairs
$\langle ij\rangle $. Although there is no long-range order, the
spins are locally ordered, with a coherence length of one lattice
spacing. The spin ordering patterns for triangular and Kagom\'{e}
lattices are summarized in Fig.\ref{tri-and-kagome}.

The r.h.s. of Eq.(\ref{u-to-S}) gives zero for the Kagom\'{e}
lattice, as in the triangular lattice, because the local spin
structures are the same in both lattices, $S_{i}\cdot S_{j}=-1/2$.
HP boson analysis of the Kagom\'{e} lattice also reveals the
absence of static lattice distortion, whether with optical or
acoustic phonons. Our findings may be summarized as follows: In
the triangular lattice where spins are long-range ordered, and the
Kagom\'{e} lattice which has a local spin ordering without the
long-range order, spin-lattice coupling fails to produce the
static lattice distortion, or the spin-Peierls effect.

Pyrochlore lattice is distinct from the previous two cases in the
sense that spins do not order even locally. Ground state manifold
of classical spins on the pyrochlore lattice is defined by the
condition $\sum_{i\in \tau} S_i = 0$ where the basic building
block $\tau$ is a tetrahedron, for all the tetrahedra forming the
lattice. The requirement is not sufficient to determine $S_i \cdot
S_j$ uniquely for nearest-neighbor sites, hence the claim that
local spin ordering is missing in the pyrochlore lattice.

Interaction energy $-(\alpha ^{2}/2)\sum_{i}f_{i}^{2}$ is minimized for $%
f_{i}^{2}$ maximum at each site. This selects the collinear spins
(all spins parallel or antiparallel to each other) as the
preferred ground state. The collinear spin patterns and the
associated lattice distortion were analyzed in details in Ref.
\onlinecite{Sondhi}, following the pioneering work on the
spin-Peierls effect in the pyrochlore structure in Ref.
\onlinecite{ueda}.

Recently, neutron scattering data on the pyrochlore compound ZnCr$_{2}$O$%
_{4} $ revealed a very interesting picture of the spin dynamics\cite{Cheong}%
. According to Ref. \onlinecite{Cheong}, each non-overlapping
hexagon embedded inside the pyrochlore lattice has six spins form
a collinear, antiferromagnetic cluster, which are energetically
decoupled from those of other hexagonal clusters. The resulting
block spins are christened \textquotedblleft spin-loop directors",
or directors for short, in Ref. \onlinecite{Cheong}. While this
picture is intuitively appealing, no quantitative justification
for the formation, and the stability of, such a hexagonal spin
cluster appears to exist to date.

In the rest of the paper we show that spin-lattice coupling can
aid in the formation of a hexagonal cluster suggested in Ref.
\onlinecite{Cheong}. We argue that each hexagon can shrink -
uniform contraction without breaking $ C_{6}$ symmetry - to
maximize the gain in exchange energy within the cluster. Spins
within each hexagon is antiferromagnetically ordered, at the
classical level. The contraction imposes a constraint, via
Eq.(\ref{u-to-S}), that the directors of nearby hexagons be
mutually orthogonal, thus lifting the huge degeneracy of the
original ground state manifold.

We first analyze the case of a single hexagonal antiferromagnetic chain
coupled to the lattice as in Eq.(\ref{spin-phonon-model}). The classical
ground state of the effective Hamiltonian is given by the staggered spin, $%
S_{i}^{0}=\pm S\hat{z}$. The staggered spins give $u_{i}$ all pointing
inward to the center of the hexagon, as given by

\begin{equation*}
\langle f_{i}\rangle =\left[ S\left( S+1\right)
-\frac{S}{6}\sum_{k}\omega _{k}\left(n_B
(k)+\frac{1}{2}\right)\right] \sum_{j\in i}\hat{e}_{ji}
\end{equation*}%
in the HP analysis. Here $\omega _{k}$ is the dispersion of a
single hexagonal unit: $\omega _{k}=2\left\vert \sin k\right\vert
$, $k=2\pi (integer)/6$. The orientation of the antiferromagnetic
spins defines the spin-loop director of Ref. \onlinecite{Cheong}.
%%%%%%%%%%%%%%%%%%%%%%%%%%%%%%%%%%%%%%%%%%%%%%%%%%%%%%%%%%%%%%%%%%%%%%%%%%%%%%%%%
\begin{figure}[b]
\includegraphics[width=8.5cm]{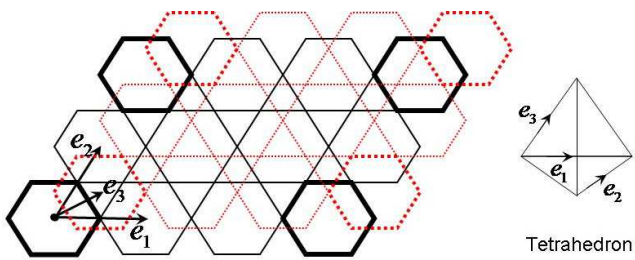}
\caption{{\protect\small Three vectors that span the pyrochlore:
$\hat{e}_1$, $\hat{e}_2$, and $\hat{e%
}_3$. $\hat{e}_1$ and $\hat{e}_2$ lie within a Kagom\'{e} plane,
while $\hat{e}_3$ connects the nearest-neighbor Kagom\'{e} planes.
Each plane is drawn by drawn by full and dashed lines. The angle
between the three vectors are shown in a single tetrahedron. Thick
black (first layer), and thick red (second layer) hexagons shrink
due to spin-lattice coupling. } } \label{Kagome}
\end{figure}
%%%%%%%%%%%%%%%%%%%%%%%%%%%%%%%%%%%%%%%%%%%%%%%%%%%%%%%%%%%%%%%%%%%%%%%%%%%%%%%%%

Taking individual hexagon as a structural unit, the pyrochlore
lattice is built up of four different types of non-overlapping
hexagons (defined A, B, C, and D), depending on the orientation of
the face of a hexagon. The four directions correspond also to the
normal of the four faces of a tetrahedron. Planes perpendicular to
each orientation defines a Kagom\'{e} lattice embedded in a
pyrochlore structure. Taking one such Kagom\'{e} plane as the
basal plane, spanned by $\hat{e}_1$ and $\hat{e}_2$, and the
vector $\hat{e}_3$ connecting two nearby basal planes, the A-type
hexagons are given the coordinates $(3m,2n,p)$, or $3m\hat{e}_1 +
2n\hat{e}_2 + p\hat{e}_3$, for integers $m,n$, and $p$
(Fig.\ref{Kagome}). The other three types are sandwiched between
two nearby Kagom\'{e} planes that are shown in Fig. \ref{Kagome}.
Coordinates of a B-type hexagon, for example, is defined by those
of the A-type lying nearest to it. The remaining hexagons are then
each located at $(B;3m,2n+1,p)$, $(C;3m-1,n,2p+1)$, and $%
(D;3m-1,n,2p)$. An $A$-type hexagon at $(3m,2n,p)$ is neighbored by four $B$%
-type hexagons located at $(3m,2n+1,p-1)$, $(3m,2n-1,p-1)$,
$(3m,2n+1,p)$ and $(3m,2n-1,p)$. Two other hexagons, of type $C$
and $D$, surround the $A$-type hexagon, too. Such ``connectivity"
of a given hexagon type to other hexagons can be worked out, for
all hexagon types.
%%%%%%%%%%%%%%%%%%%%%%%%%%%%%%%%%%%%%%%%%%%%%%%%%%%%%%%%%%%%%%%%%%%%%%%%%%%%%%%%%
\begin{figure}[t]
\includegraphics[width=6cm]{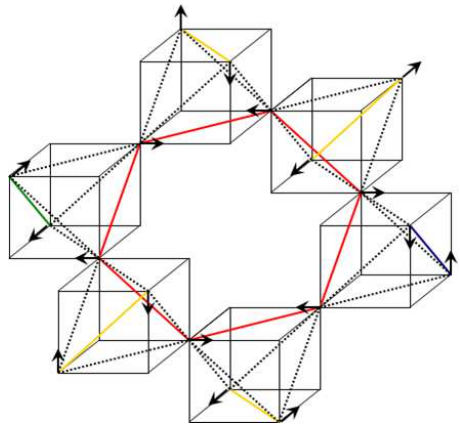}
\caption{{\protect\small Local spin configurations of the
distorted pyrochlore lattice. Short (long) bonds are denoted by
color (dashed) lines after the lattice distortion.} Each short
bond belongs to one and only one hexagon cluster.}
\label{Distortion}
\end{figure}
%%%%%%%%%%%%%%%%%%%%%%%%%%%%%%%%%%%%%%%%%%%%%%%%%%%%%%%%%%%%%%%%%%%%%%%%%%%%%%%%%

A unit cell has a $2\times 2\times 3$ structure with eight
hexagons, two of each orientation\cite{Cheong}. Each hexagon is
surrounded by six nearest-neighboring hexagons as shown in Fig.
\ref{Distortion}. According to the previous discussion, a single
hexagon undergoes a uniform contraction and has six collinear,
antiferromagnetic spins. Requiring such hexagon
contraction throughout the whole pyrochlore lattice, the r.h.s. of Eq.(\ref%
{u-to-S}) imposes the condition that the directors of the
nearest-neighbor hexagons be orthogonal
(Fig.\ref{Distortion})\cite{comment2}. A self-consistency
requirement on the hexagons is therefore that the network of
hexagons be \textquotedblleft colored" in one of three colors, say
R, G, and B, with no two neighboring hexagons having the same
color. In other words, the relative orientations of the directors
satisfy an antiferromagnetic 3-state Potts model
\begin{equation}
H_{3SP} = J_{eff}\sum_{\langle IJ\rangle }\delta (d_{I},d_{J})
\end{equation}%
for the nearest-neighbour hexagons, $\langle IJ\rangle $, and
their respective directors, $(d_{I},d_{J})$, where $d_I, d_J$
takes on R, G, or B. We find, through explicit construction, that
such coloring of the hexagon units can indeed be realized for a
pyrochlore lattice. The strength of coupling between the directors
$J_{eff} $ depends on the spin-lattice interaction $\alpha $, and
also on the level of quantum fluctuation within a hexagonal
cluster. A severe quantum fluctuation leads to the reduction of
$\langle S_{i}\cdot S_{j}\rangle $ for $i,j$ belonging to
different hexagons, and an effectively weaker $\alpha $ in
Eq.(\ref{u-to-S}). In the extreme limit $J_{eff}\rightarrow 0$ the
hexagons are completely decoupled from each other, but more
generally a residual hexagon-hexagon interaction of order
$J_{eff}$ lifts the degeneracy and leads to a band of excitation
spectra.

\begin{figure}[t]
\includegraphics[angle=90,width=8.5cm]{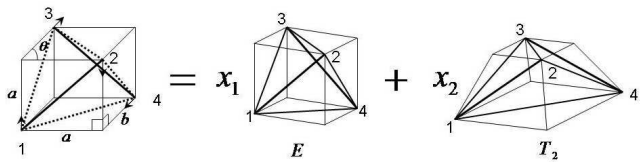}
\caption{{\protect\small Distortion of an isolated tetrahedron
consistent with a hexagon shrinkage. The spins configuration is
coplanar.}} \label{Tetrahedron}
\end{figure}

So far the distortion mode is discussed in terms of the hexagon as
a unit. Now, let us consider the distortion of an isolated
tetrahedron. For a single
tetrahedron, there are six vibrational modes: singlet $A_{1}$, a doublet $E$%
, and a triplet $T_{2}$\cite{Sondhi}. When the pyrochlore lattice assumes
the hexagonal distortion we discuss, the resulting distortion for an
isolated tetrhedron is not tetragonal ($E$) as discussed in Ref. %
\onlinecite{Sondhi}, but is a linear combination of a doublet $E$
and a triplet $T_{2}$ (Fig.\ref{Tetrahedron}).

%%%%%%%%%%%%%%%%%%%%%%%%%%%%%%%%%%%%%%%%%%%%%%%%%%%%%%%%%%%%%%%%%%%%%%%%%%%
%%%%%%%%%%%%%%%%%%%%%%%%%%%%%%%%%%%%%%%%%%%%%%%%%%%%%%%%%%%%%%%%%%%%%%%%%%%

In conclusion, lattice-coupled antiferromagnetic Heisenberg spin
model on a variety of frustrated lattices is analyzed. For a
triangular and Kagom\'{e} lattice where the ground state spin
structures are locally rigid, no spin-Peierls effect arises.
Pyrochlore spins have a continuous manifold of ground states and a
lack of local rigidity of spins, and spin-lattice coupling easily
induces a lattice distortion leading to a reduction of the ground
state degeneracy. A particular pattern of such lattice distortion,
namely a hexagon contraction, is proposed and analyzed, following
the suggestion of Ref.\onlinecite{Cheong}. Each hexagonal cluster
takes advantage of the lattice deformation to reduce its size and
maximize the exchange energy within the cluster. The hexagonal
`\emph{protectorate}' of Ref.\onlinecite{Cheong} is thus obtained.
Interaction between nearby hexagons is mediated by the
spin-lattice condition, Eq.(\ref{u-to-S}), and leads to mutually
orthogonal directors of nearby hexagons. In broader perspective,
the spin-lattice interaction provides a channel for the
self-organization of spins, which helps relieve the frustration
inherent in the underlying microscopic Hamiltonian.

\begin{acknowledgments}
We thank S.-W. Cheong and Je-Geun Park for insightful discussions. HJH is
supported by grant No. R01-2002-000-00326-0 from the Basic Research Program
of the Korea Science \& Engineering Foundation.
\end{acknowledgments}


\begin{thebibliography}{99}
\bibitem{senthil} A recent development in this direction can be found in T.
Senthil \textit{et al.} Science \textbf{303}, 1490 (2004).

\bibitem{anderson} P. W. Anderson, Science \textbf{235}, 1196 (1987).

\bibitem{theory} D. H. Lee, R. G. Caflisch, and J. D. Joannopoulos, Phys.
Rev. B \textbf{29}, 2680 (1984); Th. Jolicoeur, E. Dagotto, E.
Gagliano, and S. Bacci, Phys. Rev. B \textbf{42}, 4800 (1990); A.
B. Harris, C. Kallin, and J. Berlinsky, Phys. Rev. B \textbf{45},
2899 (1992); R. Moessner and J. T. Chalker, Phys. Rev. B,
\textbf{58}, 12049 (1998); R. Chitra \textit{et al.} Phys. Rev. B
\textbf{52}, 1061 (1995).

\bibitem{exp} H. Mamiya \textit{et al.} J. Appl. Phys. \textbf{81}, 5289; S.
H. Lee \textit{et al. } Phys. Rev. Lett. \textbf{84}, 3718 (2000);
S. H. Lee \textit{et al.} Phys. Rev. Lett. \textbf{93}, 156407
(2004); Seongsu Lee \textit{et al.} Submitted to Phys. Rev. Lett.
(2004).

\bibitem{exp-on-Jij} W. A. Harrison, \textit{Electronic Structure and the
Properties of Solids} (Dover, New York, 1980).

\bibitem{pytte} Spin-lattice interaction through modulation of the
exchange integral $J_{ij}$ was discussed in E. Pytte, Phys. Rev. B
\textbf{10}, 4637 (1974). See also references therein.

\bibitem{recent-papers} A number of papers analyzed a similar model, but
primarily on a square lattice with next-nearest-neighbour bonds. See F.
Becca and F. Mila, Phys. Rev. Lett. \textbf{89}, 037204 (2002).

\bibitem{han} June Seo Kim, Jung Ho Nam, and Jung Hoon Han, Work in progress.

\bibitem{comment} We are not concerned with a uniform expansion/contraction
of the lattice.

\bibitem{HP} S. J. Miyake, J. Phys. Soc. Japan \textbf{61}, 983 (1992); R.
R. P. Singh and D. Huse, Phys. Rev. Lett. \textbf{68}, 1766 (1992).

\bibitem{Sondhi} Oleg Tchernyshyov, R. Moessner, and S. L. Sondhi, Phys.
Rev. Lett. \textbf{88}, 067203 (2002) and Phys. Rev. B \textbf{66}, 064403
(2002).

\bibitem{ueda} Yasufumi Yamashita and Kazuo Ueda, Phys. Rev. Lett. \textbf{85%
}, 4960 (2000).

\bibitem{Cheong} S.-H. Lee, C. Broholm, W. Ratcliff, G. Gasparovic, Q.
Huang, T. H. Kim, and S.-W. Cheong, Nature, \textbf{418}, 856
(2002).

\bibitem{comment2} It can be shown that $\sum_{j\in i } (u_i - u_j ) $
is proportional to $u_i$ for a uniform contraction of hexagons in
the pyrochlore. The orthogonality condition of nearby directors
therefore applies either for optical, or acoustic phonons.

\end{thebibliography}
\end{document}